# MRSaiFE: Tissue Heating Prediction for MRI: a Feasibility Study

Simone Angela Winkler
Dept. of Radiology
Weill Cornell Medicine
New York NY USA
ORCID: 0000-0002-8080-9488

Isabelle Saniour
Dept. of Radiology
Weill Cornell Medicine
New York NY USA
ORCID: 0000-0002-0349-0311

Akshay Chaudhari
Dept. of Radiology
Stanford University
Stanford CA USA
ORCID: 0000-0002-3667-6796

Fraser Robb
Special Coils and Prototyping
GE Healthcare
Aurora OH USA
ORCID: 0000-0002-1625-9709

J Thomas Vaughan
Zuckerman Institute
Columbia University
New York NY USA

*Abstract*— **A to-date unsolved and highly limiting safety concern for Ultra High-Field (UHF) magnetic resonance imaging (MRI) is the deposition of radiofrequency (RF) power in the body, quantified by the specific absorption rate (SAR), leading to dangerous tissue heating/damage in the form of local SAR hotspots that cannot currently be measured/monitored, thereby severely limiting the applicability of the technology for clinical practice and in regulatory approval. The goal of this study has been to show proof of concept of an artificial intelligence (AI) based exam-integrated real-time MRI safety prediction software (MRSaiFE) facilitating the safe generation of 3T and 7T images by means of accurate local SAR-monitoring at sub-W/kg levels. We trained the software with a small database of image as a feasibility study and achieved successful proof of concept for both field strengths. SAR patterns were predicted with a residual root mean squared error (RSME) of <11% along with a structural similarity (SSIM) level of >84% for both field strengths (3T and 7T).**

## I. Introduction

Modern neuroscience targets the understanding of the brain in health or disease [1]. Thus, availability of technology that can significantly increase the spatial resolution and sensitivity achievable with magnetic resonance (MR) neuroimaging at emerging Ultra High-Field (UHF) of 7T or higher, consistent with safety, would offer the potential to advance our understanding of brain structure and function by enabling their investigation with greater specificity and granularity.

A key limitation to the high potential of UHF magnetic resonance imaging (MRI) in neuroscience research and clinical or diagnostic applications is safety concern related to the nonuniform deposition of radiofrequency (RF) power in the body, quantified by the specific absorption rate (SAR), which can lead to dangerous tissue heating and damage. Not only does the average SAR have a quadratic dependence on static magnetic field strength ($B_0$), increasing 4-fold from 3T to 7T, but due to the higher Larmor frequency and thus shortened in-tissue wavelength it also exhibits a spatial variation that can lead to "local SAR" patterns or "hotspots" [2]–[5] of focused high RF power deposition and localized tissue heating. Moreover, parallel transmit (pTx) technology with multiple independent transmit RF channels [6],[7] is now common in UHF applications and can lead to even stronger hotspots because of potential constructive interference of the electric fields.

While a small portion of UHF MRI has received first FDA approval for clinical routine (Siemens MAGNETOM Terra, adult head and knee imaging with one select RF coil [8]), the vast majority of clinical imaging has been performed at 3T to date. This is due to substantial safety and technological hurdles that still need to be surmounted before the potential benefits of higher sensitivity and spatial resolution can be fully realized. Specifically, there is a lack of technology that can measure local SAR due to anatomical and positional variations between patients, as well as between transmit coils. Current technology is not equipped to measure spatially varying local SAR; the only quantity that can be determined in vivo is the overall average, or global, SAR, delivered to the entire anatomy under investigation. Local SAR variation is highly difficult to predict due to anatomical and positional variations between patients, as well as transmit coil variations. Many institutions use a conservative estimate of the peak local SAR via its ratio to the measurable global SAR; typically ~20:1[9]– thereby severely limiting the applied transmit power and thus the imaging performance achievable by UHF MRI, in particular resolution and/or scan time. MR Thermometry as an alternative approach suffers from a coarse temperature resolution [10]. This critical barrier is one of the mainstay reasons why 7T has not yet reached the patient in its full capacity and holds back its success as an extremely powerful imaging modality with unprecedented ability to decipher fine structures. In this paper, we propose MRSaiFE, an artificial intelligence (AI) based exam-integrated MRI safety prediction software, facilitating the safe generation of 7T images. Using this tool, we hypothesize SAR-monitoring at sub-W/kg levels at <10%.

## II. Methods

### A. Data generation

Input data for this study was acquired from Sim4Life simulations (Zurich MedTech, Zurich, Switzerland) using the Virtual Population (IT'IS Foundation, Zurich, Switzerland).

*3T:* A 3T body coil model made for a standard bore size of 60cm was used in conjunction with the body models Duke and Ella at 224 different positions spanning from +/-60cm, +/-40cm, and +60/-100cm along the *x*-, *y*-, and *z*- axes (axial: *xy*-plane, coronal: *yz*-plane, sagittal: *xz*-plane). The anatomical input image that would come from an MRI scanner in a real experiment was approximated by using a black/white image of the voxeled body model. The 1g averaged peak local SAR output was evaluated, and coronal SAR slices were extracted (40 slices for Ella, 62 slices for Duke). This resulted in a set of 22,848 input anatomical and 22,848 output SAR images that were used to train the deep learning model, whereof 16,320 were used for training, 4,080 were used for validation, and the remainder for the test dataset.

*7T:* A smaller training set with fewer positions was used to provide first results at 7T in this proposal. A 7T birdcage head coil model was used with the body model Ella at 175 different positions within the coil ranging from +/-40 cm, +/-20cm, and 60 cm in the *x*-, *y*-, and *z*-direction (axial: *xy*-plane, coronal: *yz*-plane, sagittal: *xz*-plane). Input anatomical image and output SAR image generation followed the same steps as for the 3T analysis using 62 sagittal slices. Example figures of input anatomical image and output SAR image are shown (Table 1).

### B. Network

We implemented a unet2D architecture [11] using a cascade of convolutional filters paired with nonlinear reLU activation functions and input image sizes of 224x224 pixels. Training was performed using Adam and stochastic gradient descent (SGD) as well as Keras and Tensorflow (Google, Mountain View, California). Hyperparameters were optimized by minimizing root mean squared error (RSME) loss on the validation datasets. Training was performed over 30 epochs (SGD) and 6 epochs (Adam) using a GeForce RTX 2080 Ti Graphics Processing Unit (GPU) (NVIDIA, Santa Clara, USA). Layers were randomly initialized using the He initialization.

### C. Testing

Testing was performed on the testing datasets described in "Data generation". Quantitative image quality comparisons were performed between the ground-truth images (simulated SAR) and the predicted SAR using RMSE and structural similarity (SSIM)[12], which unlike RMSE can evaluate perceptual image quality.

## III. RESULTS

### A. Hyperparameters

Tuning of training parameters yielded optimal values of a learning rate of 0.1 with a drop of 0.1 every 15 epochs, a momentum of 0.925, batch size of 1 (3T, SGD); a learning rate of 1e-4, dropping by 0.1 every 5 epochs with $\beta_1=0.95$, $\beta_2=0.9$, and $\varepsilon=1e-4$, batch size of 1 (3T, Adam); a learning rate of 0.1 with a drop of 0.1 every 4 epochs, a momentum of 0.95, batch size of 4 (7T, SGD); a learning rate of 1e-4 with a drop of 0.1 every 5 epochs, with $\beta_1=0.8$, $\beta_2=0.995$, and $\varepsilon=1e-6$, batch size of 16 (7T, Adam).

### B. Training Time

The total duration for the training of the 3T model was 4.25 hours (SGD, residual validation loss 1.2e-3) and 50 minutes (Adam, residual validation loss 1.9e-3), while the training of the 7T model took 3.5 hours (SGD, residual loss 4.1e-3) and 40 minutes (Adam, residual loss 3.8e-3). Network convergence was observed at epoch 15-18 (SGD, 3T) and 5-6 (Adam, 3T).

TABLE I. SAR prediction results of 3T and 7T images

| | | 3T | | First results at 7T | |
|---|---|---|---|---|---|
| **Pos. 1** | Simulated data | Structural input image weighted with unloaded B1 | Simulated SAR image (1g averaged) | Structural input image weighted with unloaded B1 | Simulated SAR image (1g averaged) |
| | Prediction | Predicted SAR image (SGD optimizer) | Predicted SAR image (Adam optimizer) | Predicted SAR image (SGD optimizer) | Predicted SAR image (Adam optimizer) |
| | RMSE | 6.5% | 7.3% | 6.5% | 6.0% |
| | SSIM | 0.93 | 0.94 | 0.92 | 0.90 |
| **Pos. 2** | Simulated data | Structural input image weighted with unloaded B1 | Simulated SAR image (1g averaged) | Structural input image weighted with unloaded B1 | Simulated SAR image (1g averaged) |
| | Prediction | Predicted SAR image (SGD optimizer) | Predicted SAR image (Adam optimizer) | Predicted SAR image (SGD optimizer) | Predicted SAR image (Adam optimizer) |
| | RMSE | 6.7% | 9.9% | 10.6% | 7.2% |
| | SSIM | 0.95 | 0.86 | 0.84 | 0.88 |

*C. Image results*

Example predictions are shown in Table 1. Predicted images align well with the ground truth images. A better agreement is seen for the SGD optimizer, though Adam trains faster.

*3T results:* RMSE values were <10% for all cases with SSIM <=7% in all cases except Adam in position 2.

*7T results:* The image database was smaller to provide fast results for this proposal. Despite blurriness observed due to alignment issues between input and output data, RMSE values were found to be <=10% for all cases with SSIM <=7% in all cases except using Adam in position 2.

## IV. Discussion

The low final error for the predicted SAR despite the misalignment issues leading to relatively high residual training loss and blurry predicted images suggests robustness of the approach with respect to imperfect input data. The relatively small database used in this preliminary study in conjunction with the excellent agreement between predicted SAR and simulated SAR suggests that our large-scale SAR database will lead to small SAR errors, hopefully <5%.

Overall, MRSaiFE is expected to eventually provide UHF MRI with consistent tissue heating monitoring for use as a safe, practical, and non-invasive mainstream tool for the clinical understanding, diagnostics, monitoring, and treatment guidance at sub-mm resolution. In practice, the existing conservative SAR margins of 20:1 will be exchangeable for optimized, patient-specific margins and will free up valuable transmit power that can be used towards better sensitivity, resolution, or scan time.

This work could also significantly impact the safety of scanning patients with medical implants. Implants present a great cause for tissue heating concerns [13]. Expanding the catalog with implant patients will provide a future tool for accurate SAR prediction in these patients. This ultimately results in such patients being able to undergo MRI exams more routinely, and not only in critical situations or not at all.

The advent of even higher field strengths such as 9.4T and 10.5T[14], [15] for human MRI has brought about an even greater scrutiny and valid concern with regard to patient safety. The spatial SAR variations and average global SAR are increased compared to 7T, and MRSaiFE can be of great use in bringing these technologies to clinical practice.

In hyperthermia, tissue heating is directed at specific tissue regions with the goal of ablation [16]. UHF MRI with its intrinsic short wavelength and state-of-the-art parallel transmit capability can be used to tailor these heating hotspots by tailoring local SAR. MRSaiFE bears the potential of enabling targeted treatment planning in MR hyperthermia for cancer and other diseases in the long term.

## Conclusion

We developed proof of concept for MRSaiFE, an AI-based exam-integrated real-time MRI safety prediction software, facilitating the safe generation of UHF MRI images by means of accurate local SAR-monitoring at sub-W/kg levels. We trained the software with a small database of image as a feasibility study and achieved successful proof of concept for both field strengths. SAR patterns were predicted with a residual RSME of <11% along with an SSIM level of >84% for both field strengths (3T and 7T).

## Acknowledgments

The authors thank Michael Oberle and Erdem Ölfi at Zurich MedTech for their assistance with rapid SAR simulations.